\documentclass[conference]{IEEEtran}
\IEEEoverridecommandlockouts
\usepackage{cite}
\usepackage{multirow}
\usepackage{amsmath,amssymb,amsfonts}
\usepackage{algorithmic}
\usepackage{float}
\usepackage{graphicx}
\usepackage{textcomp}
\usepackage{xcolor}
\usepackage{url}
\usepackage{array}
\usepackage{comment}
\usepackage{siunitx}

\newcommand{\lt}{\ensuremath <}

\def\BibTeX{{\rm B\kern-.05em{\sc i\kern-.025em b}\kern-.08em
    T\kern-.1667em\lower.7ex\hbox{E}\kern-.125emX}}
\begin{document}

\title{A cryogenic readout integrated circuit with analog pile-up and in-Pixel ADC for high frame rate Skipper CCD-in-CMOS Sensors}

\author{Adam Quinn, Farah Fahim, Davide Braga


\thanks{This work was produced by Fermi Forward Discovery Group, LLC under Contract No. 89243024CSC000002 with the U.S. Department of Energy, Office of Science, Office of High Energy Physics. Publisher acknowledges the U.S. Government license to provide public access under the DOE Public Access Plan DOE Public Access Plan.

A. Quinn, D. Braga, and F. Fahim are with the Fermi National Accelerator Laboratory, Pine \& Kirk St, Batavia, IL 60510
(e-mail: \protect\url{aquinn@fnal.gov}).

Report Number: PUB-26-0540-ETD}}

\maketitle

\begin{abstract}

The Skipper CCD-in-CMOS Parallel Read-Out Circuit V2 (SPROCKET2) is an in-Pixel analog front end and ADC designed for the vertical readout of Skipper CCD-in-CMOS image sensors. SPROCKET2 is fabricated in a 65 nm CMOS process and each pixel occupies a \SI{60}{\micro\meter} $\times$ \SI{60}{\micro\meter} footprint. SPROCKET2 is intended to be heterogeneously integrated with a Skipper-in-CMOS sensor ASIC, such that one readout pixel is connected to a multiplexed array of sixteen \SI{15}{\micro\meter} $\times$ \SI{15}{\micro\meter} Skipper-in-CMOS pixels. In order to fully leverage the Skipper CCD-in-CMOS sensor's ability to achieve exceptionally low noise from repeated sampling while minimizing the power consumption of digitization and data movement, SPROCKET2 is designed to ``pile-up'' ten successive samples in the analog front end before digitizing the result with a compact 10-bit SAR ADC at a rate of 66.7 ksps, leading to an effective throughput of 667 ksps. The SPROCKET2 pixel achieves input-referred noise $ \lt 100 \mu V_{rms}$ with $\lt 1$ LSB ADC non-linearity while consuming an estimated $44 \mu W$ per pixel.
A SPROCKET2 test pixel was submitted in December 2022, and test results are presented.  

\end{abstract}

\section{Introduction}

\begin{figure}
  \centering
  \includegraphics[width=0.8\linewidth]{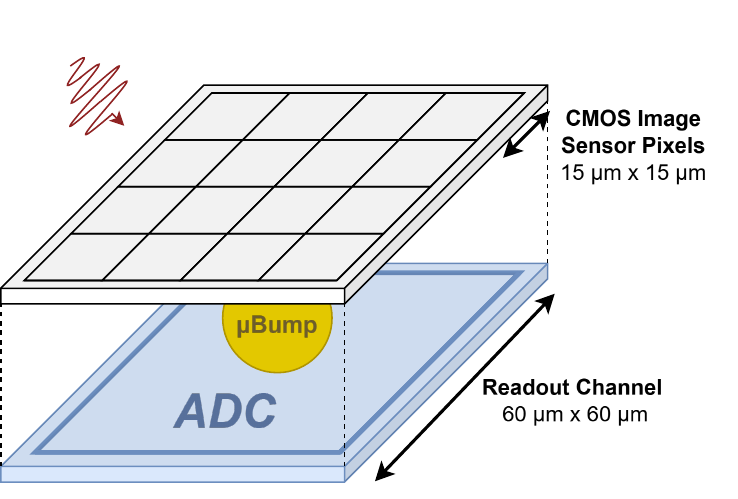}
  \caption{An illustration of the hybrid integration approach used by SPROCKET2. A single ADC pixel on SPROCKET2 is multiplexed to a $4\times4$ array of CMOS image sensor pixels.}
  \label{fig:hybrid}
\end{figure}

Future high energy physics (HEP), X-ray imaging, and quantum imaging experiments will continue to require extremely sensitive and low-noise particle detectors with large active area and high spatial resolution, leading to higher data throughput \cite{ECFA_Roadmap} \cite{kHz_Xrays} \cite{Gao2025}. Charge-Coupled Device (CCD) cameras offer excellent performance in a range of scientific imaging applications \cite{janesick1985ccd} \cite{janesick2001scientific}. In particular, Skipper-CCDs, and the Skipper-in-CMOS sensors currently being developed by Fermilab and SLAC, are a highly attractive class of detectors because they allow sub-single-electron noise to be achieved by reading out the same packet of charge many times \cite{Lapi_TED} \cite{fernandez_moroni_2012} \cite{janesick_new_1990} \cite{tiffenberg_2017}. However, traditional monolithic architectures in which charge is read out only from the edge of the CCD array cannot scale to megapixel array sizes without severely compromising readout speed \cite{Fossum_Review}. A promising alternative is offered by ``hybrid'' readout architectures which integrate a readout integrated circuit (ROIC) directly beneath the image sensor, thus allowing each Skipper CCD-in-CMOS pixel (or a multiplexed group of several pixels) to be bump bonded to a dedicated readout circuit (see Fig. \ref{fig:hybrid}).

The Skipper CCD-in-CMOS Parallel Read-Out Circuit (SPROCKET) family is a series of ROICs designed at Fermilab for this hybrid readout role. The design of in-pixel ADCs for SPROCKET is inherently challenging because each ADC must fit within a very constrained layout area and consume very little power without compromising the noise performance of the detector \cite{Wermes_2019}.

The first SPROCKET ADC design (SPROCKET1) was presented in Ref \cite{aquinn_iscas}, along with a prototype 32x64 ADC pixel array. The SPROCKET1 pixel consisted of a preamplifier, correlated double-sampling circuit, and a novel compact SAR ADC \cite{ADC_Patent}.

The SPROCKET2 ADC presented in this paper improves upon the design by introducing an analog pile-up circuit in the front end. In a typical application, Skipper CCD-in-CMOS sensors will sample the same charge packet many times and average the results to achieve single-electron and single-photon sensitivity \cite{tiffenberg_2017a}. SPROCKET2's analog pile-up circuit moves part of this averaging process into the analog domain by accumulating ten analog samples and averaging them into a single digital sample, a form of data compression at the edge. As discussed in Section \ref{sec:pileup}, this allows the ADC's effective sample rate to increase (from 100 ksps to 667 ksps) while simultaneously reducing the amount of data generated by the ADC (from 100 ksps to 66.7 ksps). In addition to the new pile-up circuit, improvements are made to the design of the SAR ADC, the pixel's power and bias distribution networks, and in-pixel reference buffering. 

Additionally, a SPROCKET2 Test Pixel was designed with built-in SAR logic which allows more extensive characterization of the ADC at higher speed than was possible with SPROCKET1.

\begin{figure*}[h!]
  \centering
  \includegraphics[width=\textwidth]{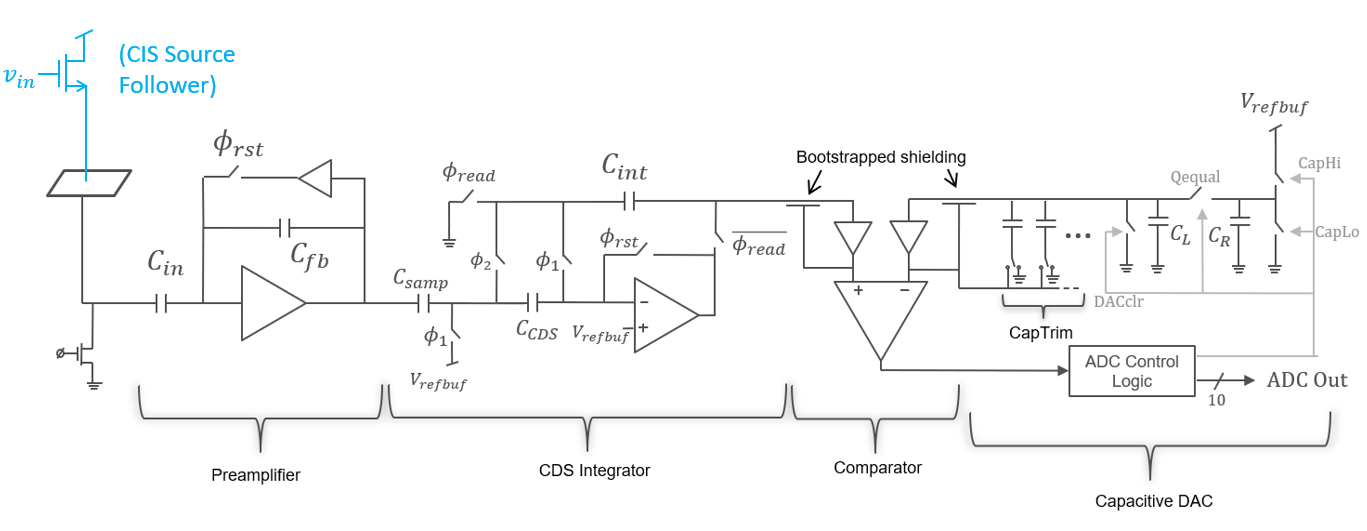}
  \caption{A simplified schematic of the SPROCKET2 ADC pixel.}
  \label{fig:sprocket_schem}
\end{figure*}

The following sections describe the design and test of SPROCKET2. Section II gives an overview of the circuit design, while Section III describes the principle of operation for in-pixel analog pile-up, which is the primary innovation of SPROCKET2. Section IV presents other design improvements implemented in SPROCKET2. Section V describes the prototype used to test this circuit, and Section VI presents results.

\section{Design Overview}

All SPROCKET ROICs are designed to operate at 100K to achieve optimal noise performance. Availability of cryogenic models is limited, and so modeling for SPROCKET2 was carried out using isothermal 77K models developed for liquid Argon experiments at Fermilab \cite{miryala2019cdp1} \cite{braga2021cryogenic}.

As shown in Fig. \ref{fig:sprocket_schem}, the input to SPROCKET2 is provided by a source-follower amplifier on the vertically-integrated Skipper CCD-in-CMOS image sensor. The source follower is biased by a current source FET on SPROCKET2, and it is followed by a capacitive-feedback preamplifier and a discrete-time CDS integrator. These blocks make up the front-end; they work to amplify and sample the input signal, and integrate multiple samples on the capacitor $C_{int}$ in preparation for read-out. 

The remainder of Fig. \ref{fig:sprocket_schem} shows the in-pixel SAR ADC, which is substantially similar to the design presented in \cite{aquinn_iscas}. A capacitive DAC (CDAC) generates binary-weighted approximation voltages by charge-sharing between two capacitors, $C_{R}$ and $C_{L}$, which are trimmed to match each other. The result is compared to the voltage stored across $C_{int}$, and the output of the comparator determines the next bit in the successive approximation. 

Analog components of the SPROCKET2 pixel operate from a 2.5V supply. Constant-current in-pixel level shifters (not shown) ensure that 2.5V control signals can be controlled by digital circuits operating from a 1.2V core supply without introducing unnecessary power supply noise.

\section{In-Pixel Analog Pile-up}
\label{sec:pileup}

SPROCKET2 performs in-pixel analog pile-up of samples to increase sample rate within a limited power budget. This pile-up occurs in three phases. During the Reset phase ($0.5\mu s$), charge stored in the front end is cleared. During the Sampling phase ($(0.5\times N) \mu s$), $N$ samples are taken and piled up on $C_{int}$. Finally, during the Conversion phase ($9.5 \mu s$), the voltage on $C_{int}$ is digitized. In the default mode, ten samples are accumulated and the entire process takes $15 \mu s$, yielding a data rate of $[10 \text{ samples}] / [15 \mu s] = 667\text{ ksps}$. The following sections describe each phase in detail, along with the novel two-step re-referencing technique that was developed to transfer charge from the integrator to the ADC between the Sampling and Conversion phases.

\subsection{Reset Phase}

In the reset phase, $\phi_{rst}$ and $\phi_1$ are asserted. The preamplifier and the integrator are reset, and the voltages across $C_{int}$ and $C_{samp}$ are cleared. The negative input of the integrator settles to $V_{ref,buf}+V_{OS}$ where $V_{OS}$ is the offset voltage of the amplifier. Thus $V_{OS}$ is stored across $C_{CDS}$, allowing it to be cancelled later. 

During the reset phase, a special control pattern is executed on the CDAC to pre-charge the negative input of the comparator to 750 mV, which will be necessary for the autoranging logic to function later.

\subsection{Sampling Phase}

\begin{figure}
  \centering
  \includegraphics[width=\linewidth]{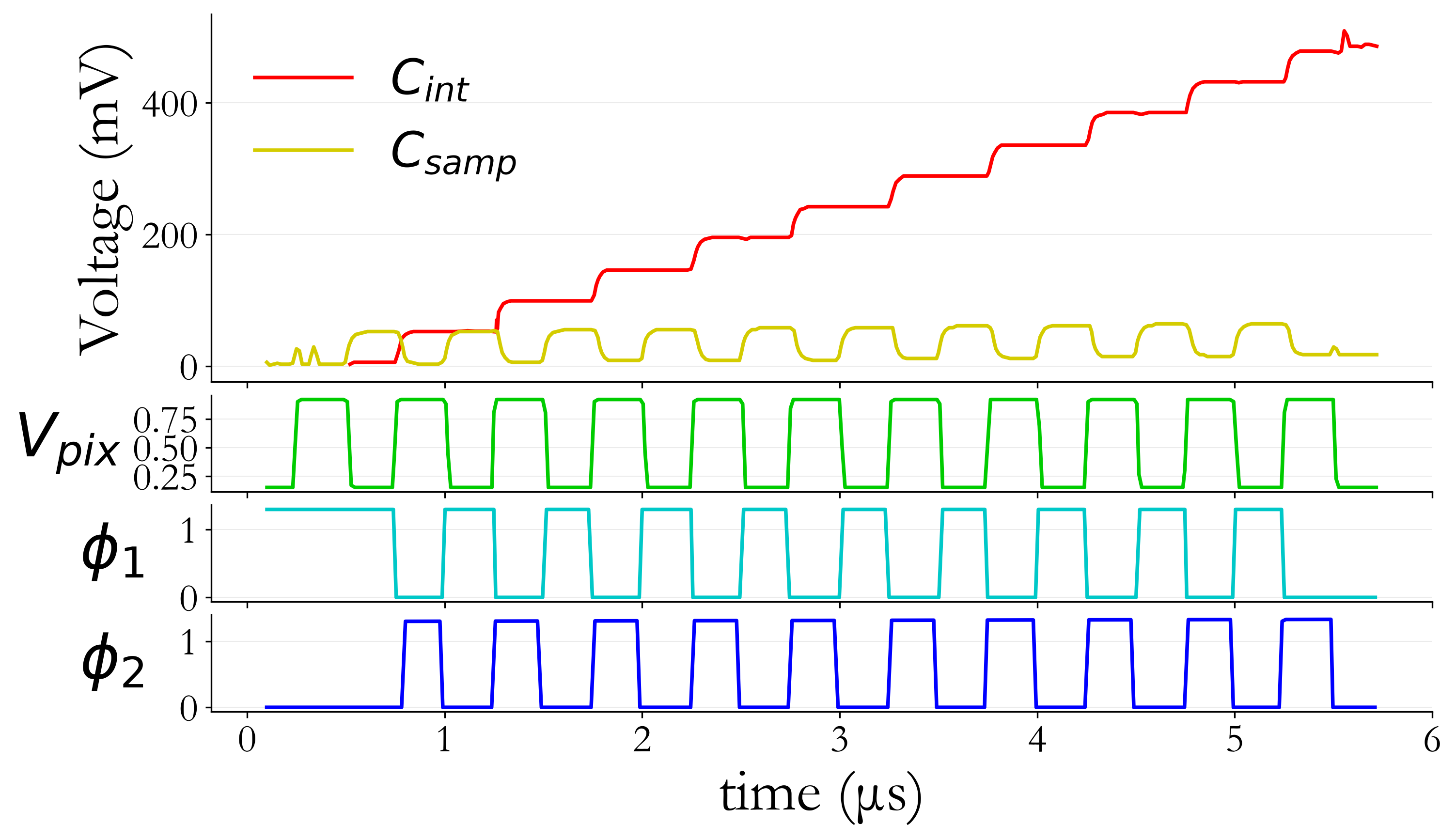}
  \caption{A simulation of the Sampling Phase. $C_{int}$ and $C_{samp}$ are the voltages across the integration and sampling capacitors respectively. $V_{pix}$ is the signal generated by the Skipper CCD-in-CMOS sensor. (This signal is typically on the order of $\mu V$, and it is scaled here for clarity.) $\phi_1$ and $\phi_2$ are non-overlapping clocks. During the Sampling Phase, each successive sample is first stored on $C_{samp}$ and then transferred to $C_{int}$.}
  \label{fig:sampling_simulation}
\end{figure}

The sampling phase operates on a non-overlapping two-phase clock ($\phi_1$,$\phi_2$) which is synchronized with the operation of the Skipper-CCD as shown in Fig \ref{fig:sampling_simulation}.

In the first phase, $\phi_1$ is asserted, and the Skipper CCD-in-CMOS drives the signal to be measured, such that the voltage at the input of SPROCKET2 is $V_{reset,ccd}-V_{sig}$. The preamplifier amplifies the signal by a factor of  $A=-\frac{C_{in}}{C_{fb}}$. A range switch (not shown) sets the value of $C_{in}$ to select either $1\times$ or $10\times$ gain. Thus a voltage of $AV_{sig}$ appears across $C_{samp}$.

In the second phase, $\phi_2$ is asserted, and the Skipper CCD-in-CMOS is reset, such that the voltage at the input of SPROCKET2 is $V_{reset,ccd}$. The preamplifier forces the left terminal of $C_{samp}$ to return to $V_{ref,buf}$. However, because the integrator's feedback path is now connected, it also forces the right terminal of $C_{samp}$ to equal $V_{ref,buf}$, which results in the charge shifting to $C_{int}$. Neglecting any charge loss, the voltage across $C_{int}$ is now $\gamma A V_{sig}$ where $\gamma$ is the ratio $\frac{C_{samp}}{C_{int}}$ which is nominally unity. 

\begin{figure}
  \centering
  \includegraphics[width=\linewidth]{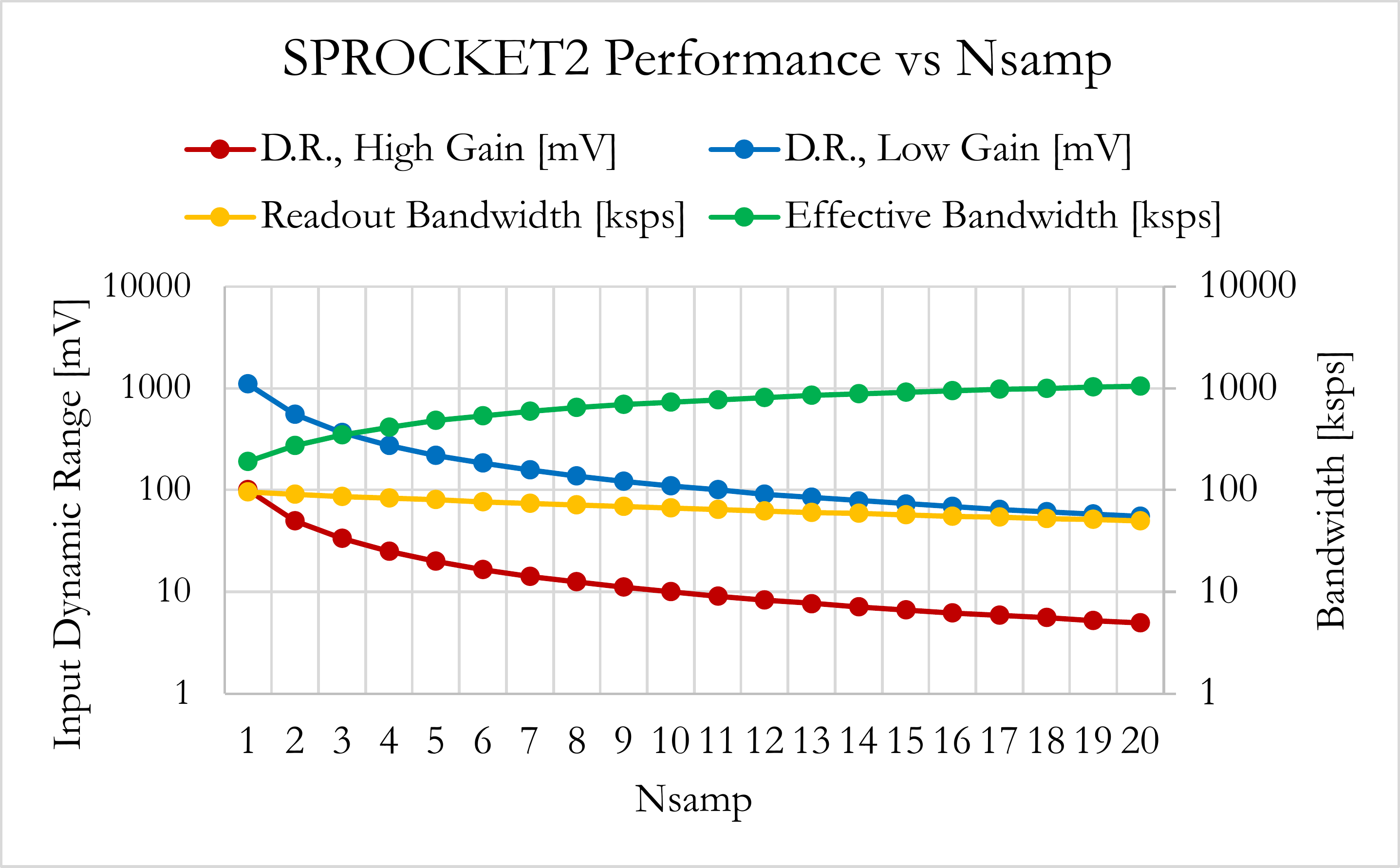}
  \caption{Impact of the number of samples piled-up before digitization ($N$) on per-sample dynamic range when the preamplifier is in high-gain mode (red), per-sample dynamic range when the preamplifier is in low gain mode (blue), the number of digital samples taken per second (yellow), and the effective number of analog samples taken per second (green). When simulating and testing SPROCKET2, we chose $N=10$.}
  \label{fig:Nsamp_Tradeoffs}
\end{figure}

This process can continue for an arbitrary number of samples. At each cycle, the new sample acquired on $C_{samp}$ is ``piled-up'' on top of the existing samples on $C_{int}$. The optimal number of samples depends on experimental needs: taking a larger number of samples before digitization results in a larger effective throughput and greater power efficiency because ADC conversion time and power is amortized over the number of samples. However, the overall dynamic range of the ADC is limited to 1.0V, so accumulating a larger number of samples reduces the effective dynamic range per sample. This trade-off is illustrated in  Figure \ref{fig:Nsamp_Tradeoffs}.

In SPROCKET2, digital autoranging logic implemented in each pixel autonomously decides whether to take one or ten samples. These numbers were chosen as a compromise between bandwidth and dynamic range: By changing the digital logic, the same analog circuits can be used to take more or fewer samples. After the first (and only the first) cycle, an autorange check triggers, which records the output of the comparator. The comparator's negative input was pre-charged to 750 mV in the Reset Phase, and its positive input is connected to the right terminal of $C_{int}$ which has a voltage of $V_{ref,buf}+\gamma A V_{sig} = 650 \mathrm{mV} + \gamma A V_{sig}$, thus the comparator output is logic high if and only if the sampled input voltage is greater than 100 mV. If this is the case, then the autorange logic will immediately stop sampling, as accumulating ten samples with a magnitude $\geq$ 100 mV is guaranteed to overflow the ADC dynamic range of 1.0V. If the comparator output is logic low, sampling will continue. 

\subsection{Two-Step Re-referencing Technique}

At the end of the Sampling Phase, the voltage across $C_{int}$ is $N\gamma A V_{sig}$ where $N$ is the number of times the input is sampled. However, $C_{int}$ is offset from ground by $V_{ref,buf}= 650 \mathrm{mV}$, which is greater than one half of the dynamic range of the ADC. In order to avoid digitizing this large offset, we must ``re-reference'' the sampled voltage from $V_{ref,buf}$ to ground. We developed a two-step technique to do so while avoiding nonlinear signal-dependent charge injection. 

To the best of our knowledge, this technique is a novel contribution. It is useful any time the output of a switched-capacitor circuit must be re-referenced, so for the benefit of future readers, we present a generic treatment of the technique as applied to a conventional switched capacitor amplifier.

\begin{figure}
    \centering
    \includegraphics[width=\linewidth]{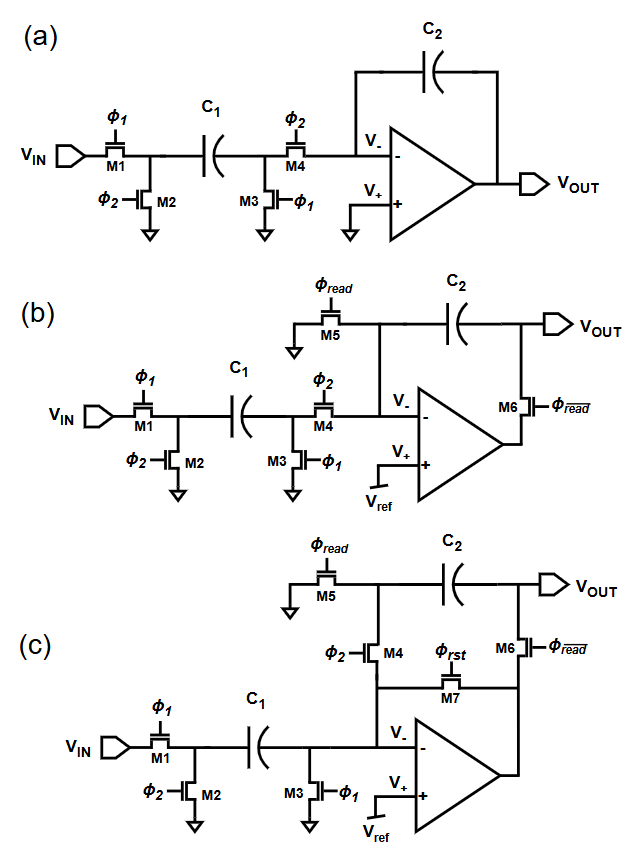}
    \caption{(a) A conventional switched-capacitor amplifier with an implicit dual supply. (b) A single-supply switched-capacitor amplifier with a naive approach for re-referencing its output voltage to ground. (c) A single-supply switched-capacitor amplifier which uses the proposed re-referencing technique.}
    \label{fig:tsrr_circuits}
\end{figure}

A conventional switched-capacitor amplifier consists of an input sampling capacitor $C_1$, a feedback capacitor $C_2$, and a differential amplifier as shown in Fig. \ref{fig:tsrr_circuits}(a). In the first phase, the input signal is sampled onto $C_1$. In the second phase, the $V_{in}$ is grounded and the amplifier's $V_{-}$ is driven to virtual ground, causing charge to redistribute from $C_1$ to $C_2$ and create a voltage of $V_{out}=-V_{in}\frac{C_1}{C_2}$.

In this conventional circuit, the differential amplifier's positive input $V_{+}$ is connected to ground, which implies the use of a dual-supply amplifier. However, in many applications (including SPROCKET2), the additional power consumption and complexity of a dual-supply amplifier is undesirable. If the differential amplifier operates from a single supply, its positive input $V_{+}$ must be connected to a reference voltage $V_{ref}$ between ground and the supply voltage, resulting in an output of $V_{out}=V_{ref}-V_{in}\frac{C_1}{C_2}$.

In this configuration, the output is a floating voltage referenced to $V_{ref}$. If the input of the stage following the switched-capacitor amplifier is referenced to ground or a different reference voltage, then the voltage on $C_2$ must be re-referenced.

Figure \ref{fig:tsrr_circuits}(b) shows a naive approach to re-referencing the output of the switched-capacitor amplifier. When $\phi_{read}$ is asserted, $C_2$ is disconnected from the output of the amplifier and its left terminal is grounded, yielding a voltage of $V_{out}\approx-V_{in}\frac{C_1}{C_2}$ on the right terminal. For simplicity, we assume that the subsequent stage is referenced to ground. In this case, $V_{in}$ must be negative with respect to $V_{ref}$ to ensure a positive output. However, in the general case the output may be re-referenced to any voltage, in which case this constraint does not hold. 

Unfortunately, the approach in Figure \ref{fig:tsrr_circuits}(b) will result in signal-dependent charge injection onto the output node via M6 because the source and drain of M6 are at the potential $V_{out}=V_{ref}-V_{in}\frac{C_1}{C_2}$ when $\phi_{read}$ is asserted. Note that signal-dependent charge injection may be completely avoided during the $\phi_1$ and $\phi_2$ phases through the use of bottom plate sampling. The technique we will present can accomplish re-referencing while avoiding signal-dependent charge injection in the $\phi_{read}$ phase as well. 

Compared to the naive approach (Figure \ref{fig:tsrr_circuits}(b)), the two-step technique  requires only one additional switch, M7, as shown in Figure \ref{fig:tsrr_circuits}(c). Additionally, M4 is moved so that it separates $V_{-}$ from $C_2$ instead of $C_1$.

When the amplification phase is complete, both $\phi_1$ and $\phi_2$ are de-asserted. At this point, re-referencing is accomplished in two steps: 

First, $\phi_{rst}$ is asserted. The differential amplifier now acts as a unity gain buffer, driving $V_{out}$ to $V_{ref}$. 

Second, $\phi_{read}$ is asserted. As in the naive approach, $C_2$ is disconnected from the amplifier and its left terminal is grounded, resulting in  $V_{out}\approx-V_{in}\frac{C_1}{C_2}$. However, because the source and drain of M6 are always charged to the potential $V_{ref}$ when it is switched off, only a constant offset is added to the output signal.

Parasitics at both terminals of $C_2$ will cause an error in the final result. However, this error is linear and thus may be viewed as a correction to the gain of the switched-capacitor amplifier. If the parasitic capacitances to ground at the left and right terminals of $C_2$ are $C_{pL}$ and $C_{pR}$ respectively, the final corrected gain of the amplifier is:

$$V_{out} = -V_{in}\frac{C_2}{C_1}(\frac{C_2}{C_2 + C_{pL}})(\frac{C_2}{C_2 + C_{pR}})$$

Our implementation in SPROCKET2 directly follows from this general treatment, replacing the switched-capacitor amplifier with our CDS integrator.

\subsection{Conversion Phase}
 
The remainder of the Conversion Phase proceeds as described in Ref \cite{aquinn_iscas}. the ADC generates a reference voltage by briefly asserting CapHi or CapLo to charge $C_R$ either to $V_{ref}$ or to zero volts, and then asserting Qequal to short the positive terminals of $C_R$ and $C_L$ together \cite{ADC_Patent}. The final voltage developed at the positive terminal of $C_L$ (the negative input to the comparator) after $N$ phases is given by:

$$\sum_{n=1}^{N} \frac{V_{ref} h_n}{2^{N-n+1}}$$

Where $h_n$ is 1 if CapHi is asserted on clock cycle $n$ and 0 if CapLo is asserted. As this expression suggests, the DAC requires $n$ clock cycles to compute an $n$-bit approximation. 

After each approximation is computed, the ADC compares it to the sampled voltage stored across $C_{samp}$ and $C_{pre}$ to decide the next $h_n$ and thus the value of the next approximation, which will have a resolution one bit greater than the previous. DACclr is then asserted to clear the value on $C_L$. Thus one 10-bit ADC acquisition requires $\sum_{n=1}^{10} n = 55$ clock cycles to compute 10 approximations of increasing precision. 

\section{Other Design Improvements}

\subsection{In-Pixel Reference Buffering}

Just as in SPROCKET1, the full scale range of the ADC is set by a reference voltage $V_{ref,adc}=1.0 \mathrm{V}$. However, the discrete time integrator requires a second reference voltage of $V_{ref,fe}=650\mathrm{mV}$, which is less than 1.0V to preserve adequate headroom. Both references must be low impedance, which suggests the use of in-pixel buffering due to limited metal routing resources in a large pixel matrix.

To avoid the power and area costs of adding a second buffer to each pixel, we recognized that $V_{ref,fe}$ is only required during the Sampling phase and $V_{ref,adc}$ is only required during the Conversion phase, allowing us to implement time-sharing of a single buffer. A periodic digital control signal $bufsel$ controls which reference is connected to the buffer's input (not shown), and the buffer's output $V_{ref,buf}$ is connected to both the ADC and the front end as shown in Figure \ref{fig:sprocket_schem}.

\subsection{CDAC Capacitor Layout}

\begin{figure}
  \centering
  \includegraphics[width=0.8\linewidth]{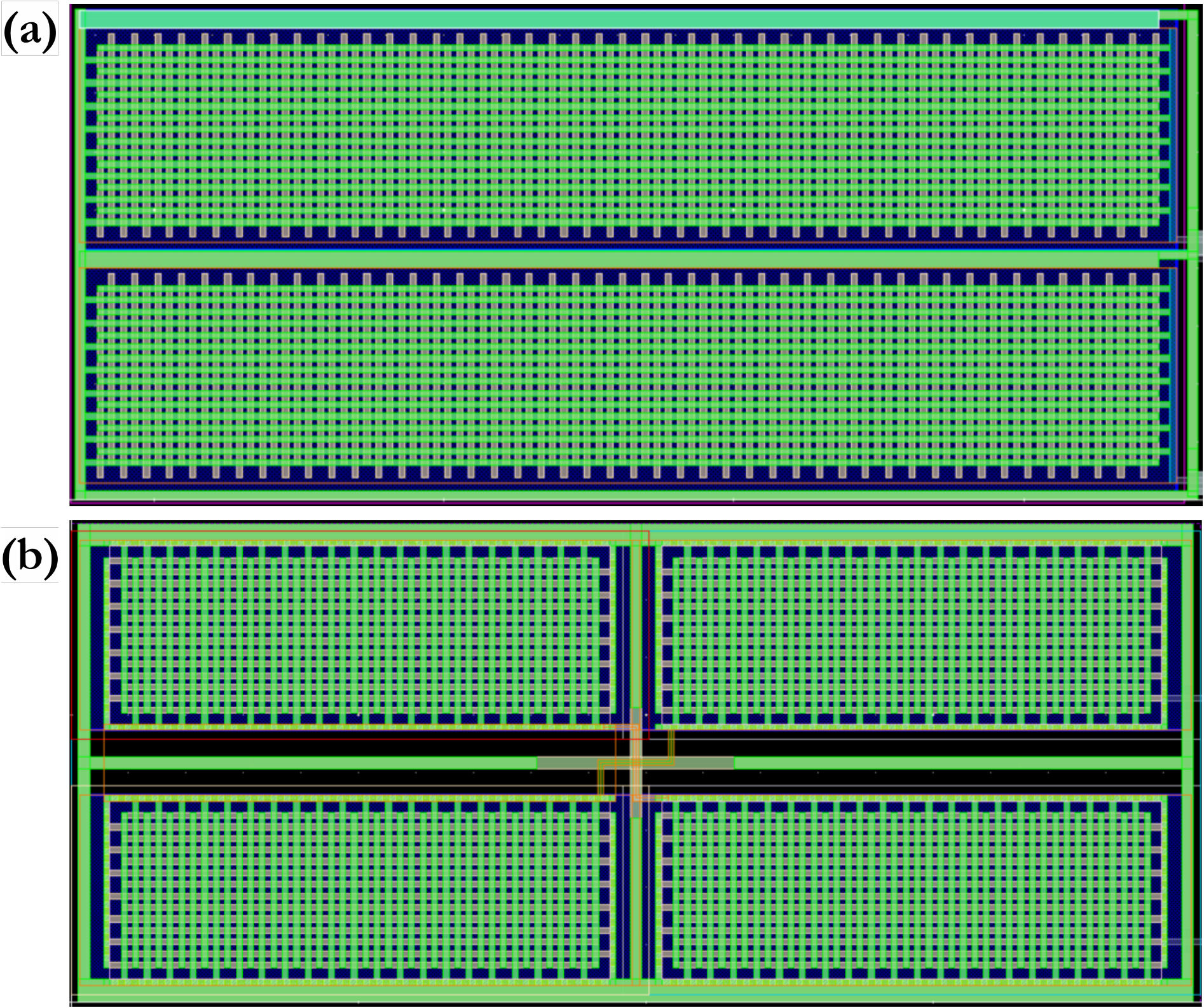}
  \caption{The pair of matched capacitors $C_{L}$ and $C_{R}$ which form the core of the capacitive DAC. (a) SPROCKET1 layout (b) SPROCKET2 layout.}
  \label{fig:cappair}
\end{figure}

Test results from SPROCKET1 showed that the trimming range of $C_L$ and $C_R$ was insufficient to compensate process mismatch in the worst case. SPROCKET2 improves upon the capacitor design by using a cross-coupled unit cell layout, shown in Figure \ref{fig:cappair}, which provides better inherent matching.

\section{SPROCKET2 Test Pixel}

\begin{figure}
  \centering
  \includegraphics[width=\linewidth]{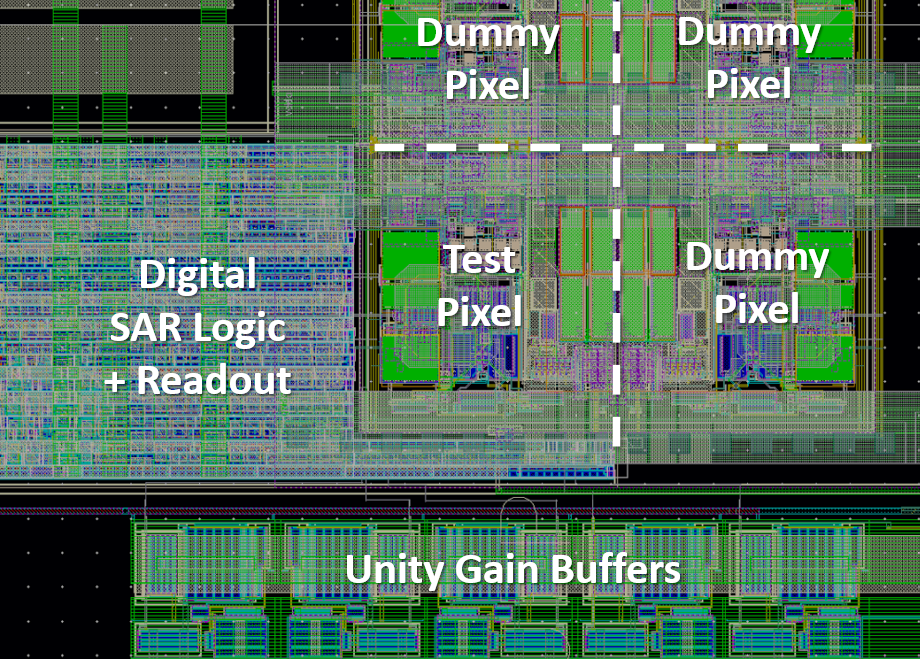}
  \caption{Layout capture of the SPROCKET2 Test Pixel and surrounding circuitry.}
  \label{fig:test_pix_layout}
\end{figure}

In order to characterize the performance of the SPROCKET2 ADC, a single pixel test structure was designed and fabricated. In this test structure, shown in Figure \ref{fig:test_pix_layout}, the single pixel under test is abutted to three dummy pixels to form a square ``island'' of custom circuitry surrounded by digital standard cells, which closely mimics the layout surroundings of the pixel when integrated into a full-chip array. Several internal analog signals from the pixel are connected to unity gain buffers which drive those signals onto pads for test observability, and a synthesized digital block is implemented next to the pixel which contains control logic for the SAR ADC as well as a scan chain for reading out the ADC result. 

The test pixel has two limitations which affect the data we collected: 

First, due to an RTL bug, the scan chain in this test chip is inoperable. Fortunately, the ADC code can also be inferred algorithmically from measuring the CapLo waveform generated by the on-pixel SAR logic. All results in the following Test Results section are captured using this alternate method.

Second, the SPROCKET2 test pixel is taped out on a shared die with a different ASIC prototype. At the top level, SPROCKET2's power rails are shared with the other ASIC, which consumes much larger power and cannot be disabled. As a result, it is not possible to make an accurate power measurement of the SPROCKET2 test pixel alone. However, the test pixel is designed to consume a constant current which can be derived from external current biases and current mirror ratios to be $44 \mu W$. We expect that the pixel consumes $\approx 44 \mu W$ at both room temperature and cryogenic temperature, with variation dominated by current mirror mismatch.


\section{Test Results}

The SPROCKET2 test pixel was evaluated at room temperature and at a temperature of 100K using a closed-cycle cryocooler at Fermilab.

\subsection{ADC Non-Linearity}

\begin{figure*}
  \centering
  \includegraphics[width=\textwidth]{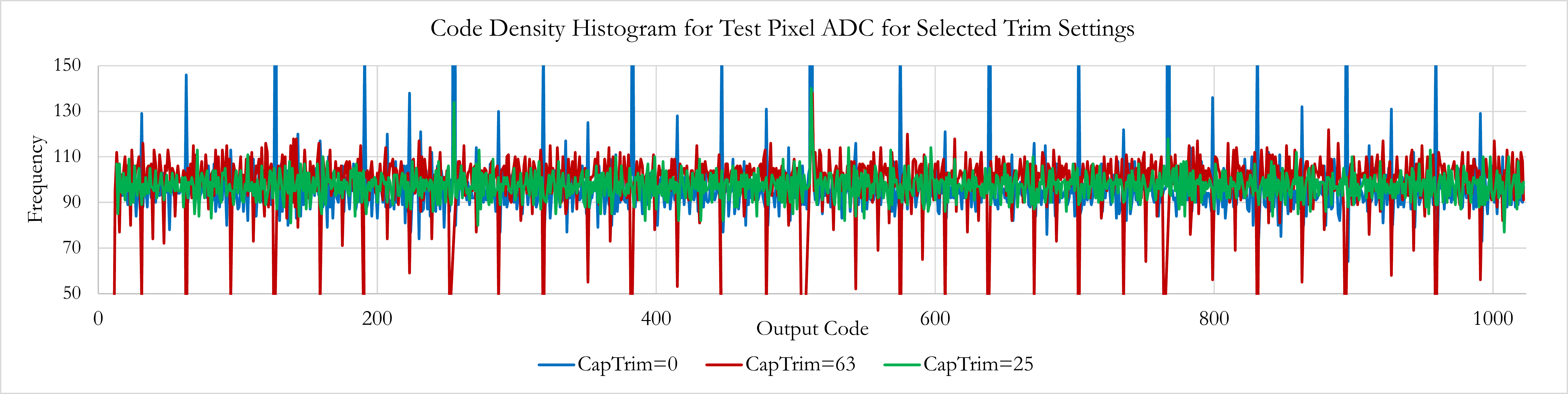}
  \caption{Code Density Histogram for the SPROCKET2 ADC}
  \label{fig:code_density_histogram}
\end{figure*}

The test pixel ADC was evaluated independently using the $V_{test}$ input (not shown), which allows a voltage to be applied directly to the input of the ADC (the positive input of the comparator), bypassing the front end.

A code density histogram approach was used to characterize the non-linearity of the ADC for various trim capacitor settings. Results for both endpoints of the trim range, as well as the optimal code for the tested device (CapTrim=25) are presented in Figure \ref{fig:code_density_histogram} and Table \ref{tab:adc_linearity}. Testing of a large number of ADCs will be required to evaluate the true range of manufacturing variation, but the fact that the test pixel lies near the center of the trim range (an ideal ADC would have optimal CapTrim=31) is a good sign.

\begin{table}[htbp]
\renewcommand{\arraystretch}{1.3}
\centering  
\caption{ADC Non-Linearity Across Range of Trim Codes (Optimal Non-Linearity Achieved with CapTrim=25)}
\label{tab:adc_linearity}
\begin{tabular}{lccc}
\hline
\hline
\textsc{CapTrim = } &  \textsc{0 (min)} &  \textsc{25 (opt)} &  \textsc{63 (max)}\\
\hline
Worst DNL [LSB] &  $5.95$ &  $0.44$ & $-1$ (missing codes)\\
Worst INL [LSB] & $3.30$ & $0.58$ & $1.38$\\
\hline
\hline
\end{tabular}
\end{table}

The presented results are measured with the ADC clock operating at 600 kHz, which is 1/10th of its design speed. Regrettably, unexpected parasitic effects from the unity gain buffers resulted in a larger than expected delay in the operation of the test pixel's comparator, which prevents it from delivering accurate results at the design speed. Because the unity gain buffers are present only in the test pixel, this issue does not affect the viability of the core SPROCKET2 circuits.

\subsection{Front End Gain}

The front end of SPROCKET2 has four distinct gain regions depending on whether the preamplifier is configured for high gain or low gain and whether the autorange function limits pileup to one sample or ten. Separately, the regions are:

\begin{itemize}
    \item \textbf{Region 0:} High-gain preamplifier, 10 samples (nominal gain: 100)
    \item \textbf{Region 1:} High-gain preamplifier, 1 sample (nominal gain: 10)
    \item \textbf{Region 2:} Low-gain preamplifier, 10 samples (nominal gain: 10)
    \item \textbf{Region 3:} Low-gain preamplifier, 1 sample (nominal gain: 1)
\end{itemize}

At both room temperature and cryogenic temperature, gain is within 5\% of the nominal value, as shown in Table \ref{tab:adc_gain}.

\begin{table}[!t]
\renewcommand{\arraystretch}{1}
\centering    
\caption{Front End Gain Results by Region}
\label{tab:adc_gain}
\begin{tabular}{lccc}
\hline
\hline
\textsc{Region} &  \textsc{Design} &  \textsc{Measured (RT)} &  \textsc{Measured (100K)}\\
\hline
0 &  100 &  98.04 &  96.00 \\
1 & 10 & 9.54 & 9.88 \\
2 & 10 & 9.69 & 10.05 \\
3 & 1 & 1.02 & 1.03 \\
\hline
\hline
\end{tabular}
\end{table}

\begin{table}[!t]
\renewcommand{\arraystretch}{1}
\centering    
\caption{Noise Results by Region}
\label{tab:noise}
\begin{tabular}{lccc}
\hline
\hline
\textsc{Region} &   \textsc{Room Temperature [uVrms]} &  \textsc{100K [uVrms]}\\
\hline
0 &  60.6 & 89.8* \\
1 &  202 & 200.6 \\
2 &  653.5 & 650.5 \\
3 &  2000 & 1500 \\
\hline
\hline
\end{tabular}
\end{table}

\subsection{Noise}

\begin{figure}
  \centering
  \includegraphics[width=\linewidth]{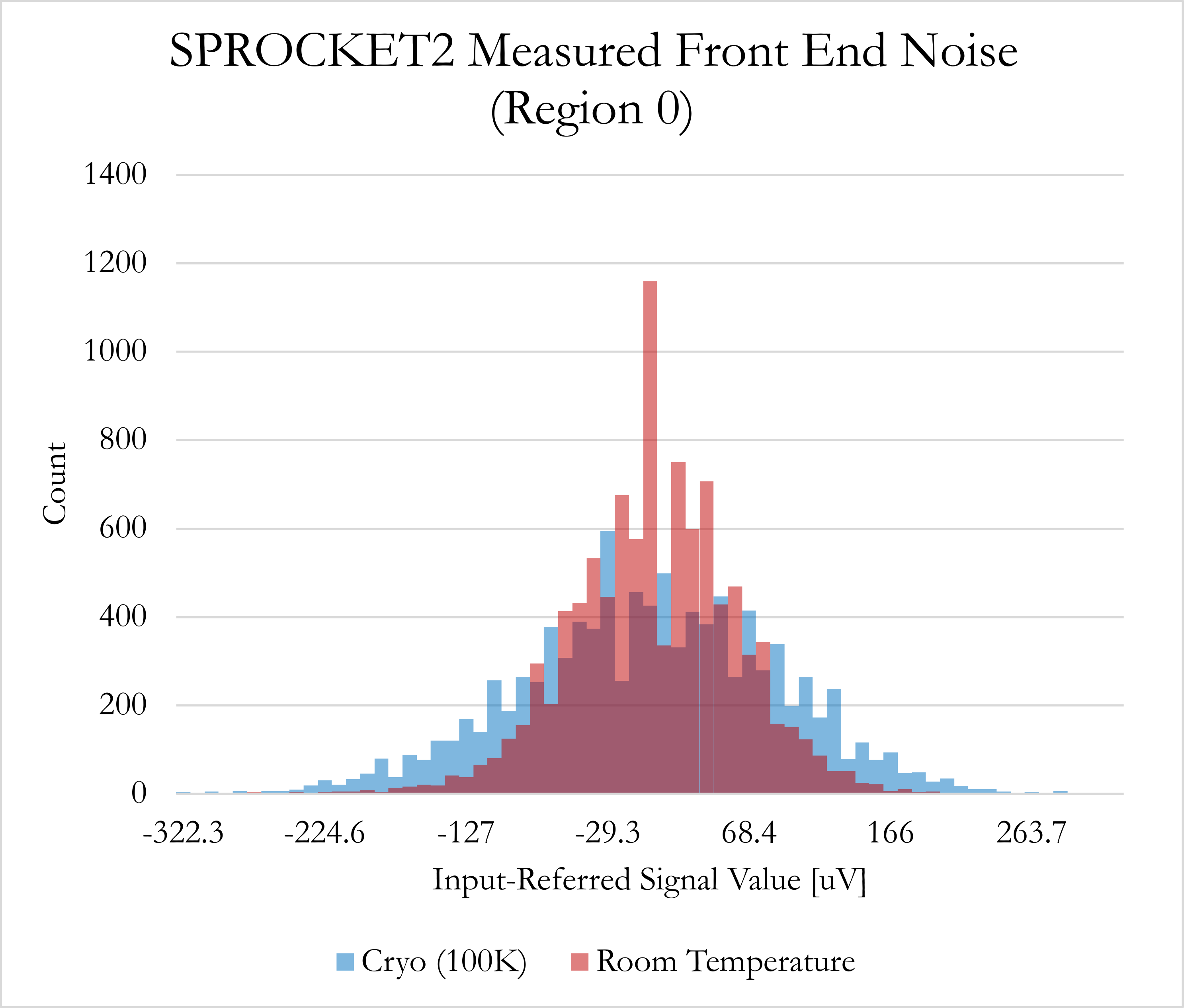}
  \caption{Histogram of measured noise results for the SPROCKET2 pixel operating in Region 0 (room and cryogenic temperature).}
  \label{fig:noise_histograms}
\end{figure}

We measured noise of the full SPROCKET2 pixel by applying a fixed input voltage, constructing a histogram of the resulting codes, and fitting a Gaussian function to the distribution. Fig \ref{fig:noise_histograms} shows the noise histograms for Region 0, the lowest-noise region, at room temperature and cryogenic temperature. Results for all four regions are reported in Table \ref{tab:noise}.

Note that Region 3 and Region 1 show an input-referred noise amplitude approximately $\sqrt{10}\approx 3.2$ times higher than Region 2 or Region 0 respectively, which is in agreement with the expected $\sqrt{N}$ improvement from accumulating multiple samples. 

At room temperature, Region 3 and Region 2 show an input-referred noise amplitude approximately $10\times$ higher than Region 1 or Region 0 respectively. Because the difference between these regions is the gain setting of the preamplifier, this suggests that the noise is dominated by internal circuit noise occuring after the preamplifier. By contrast, the measured noise for Region 0 at cryogenic temperature deviates from this pattern, suggesting that there is significant input noise, possibly due to increased environmental pickup from long cables used for cryogenic testing. We believe this is the reason that cryogenic noise exceeds room temperature noise in Region 0, and thus we report cryogenic noise for Region 0 with an asterisk. The expected value for this region based on scaling is $\approx 60 \mu V_{rms}$.

\section{Conclusions and Future Work}
We demonstrated a compact, low-power readout architecture for highly-parallel, high-rate hybrid readout of sub-electron noise Skipper-in-CMOS image sensors. The SPROCKET2 ADC architecture provides an unique solution for effective frame rates approaching 1Mfps, and was validated by measured results at a cryogenic temperature of 100K.  

The next-generation SPROCKET chip (SPROCKET3) has already been taped out in December 2023. This chip integrates more than 20,000 ADC pixels using the architecture described above, along with on-chip biasing, pattern generation, and high-speed, low-power readout using silicon photonics. Ultimately the SPROCKET3-FR (SPROCKET3 Full Reticle Version) will expand the SPROCKET read-out architecture to encompass a full 65nm reticle.



\bibliographystyle{IEEEtran}
\bibliography{IEEEabrv,biblio}
\end{document}